\documentclass[12pt]{iopart}

\usepackage{amssymb}
\usepackage{graphicx}
\usepackage[usenames,dvipsnames]{color} 
\usepackage[usenames,dvipsnames]{xcolor} 
\newcommand{\beq}{\begin{equation}}
\newcommand{\eeq}{\end{equation}}
\newcommand{\bea}{\begin{eqnarray}}
\newcommand{\eea}{\end{eqnarray}}
\newcommand{\eps}{\varepsilon}

\begin{document}

\title{The first self-consistent calculation  of quadrupole
moments of odd semi-magic nuclei accounting for phonon induced corrections}

\author{E E  Saperstein$^{1,2}$, S Kamerdzhiev$^{1}$, D S  Krepish$^{1,3}$,
S V  Tolokonnikov$^{1,3}$, and D Voitenkov$^{4}$}
  \address{$^1$National Research Centre ``Kurchatov Institute'', 123182 Moscow, Russia}
  \address{$^2$National Research Nuclear University MEPhI, 115409 Moscow, Russia}
  \address{$^3$Moscow Institute of Physics and Technology, 141700 Dolgoprudny, Russia}
 \address{$^4$Institute for Physics and Power Engineering, 249033 Obninsk, Russia}

\hskip 1.7 cm E-mail: saper43\_7@mail.ru, kaevster@gmail.com, krepish@phystech.edu,\\
.\hskip 2.5 cm tolkn@mail.ru, voitenkov@list.ru

\begin{abstract} The self-consistent model, developed previously  to
describe phonon coupling (PC) effects in magnetic moments of odd magic and semi-magic nuclei, is
extended to quadrupole moments. It is based on the theory of finite Fermi systems with the use of the
perturbation theory in $g_L^2$, where $g_L$ is the vertex creating the $L$-phonon. Accounting for the
phonon tadpole diagrams is an important ingredient of this model. The calculation scheme is based on
the Fayans energy density functional DF3-a and does not contain any adjusted parameters. The odd In
and Sb isotopes are considered, which are the proton-odd neighbors of even tin nuclei. The $2^+_1$
phonon is taken into account which quadrupole moment is one  ingredient  of the calculation scheme.
The corresponding values were found by us previously. Two main PC corrections, due to the phonon
$Z$-factor and due to the phonon-induced interaction, have opposite signs and cancel strongly each
other, leaving room for other `small' corrections, so that the resulting PC correction is much lower
than the absolute values of each of two main ones. However, it remains noticeable, making the overall
agreement with the data significantly better.
\end{abstract}

\maketitle

\section{Introduction}

In the last decade, a lot of new data on the electromagnetic moments of ground states of odd nuclei
appeared, due to intensive working of modern Radioactive Ion Beam facilities, in combination with new
spectroscopy techniques using high-intensity lasers. As a result, the bulk of the data on nuclear
static moments becomes very extensive and comprehensive \cite{Stone-2014} creating a challenge to
nuclear theory. For nuclear magnetic moments, this challenge was partially responded recently within
the self-consistent theory of finite Fermi systems (TFFS) \cite{scTFFS}, first at the mean-field level
\cite{mu1,mu2} and then with account for the particle-phonon coupling (PC) effects
\cite{EPL_mu,Lett_mu,YAF_mu}. More recently, the first self-consistent mean-field calculations of the
quadrupole moments were carried out in \cite{BE2,QEPJ,QEPJ-Web}. In this work, we find for the first
time the PC corrections to the mean-field predictions of the self-consistent TFFS for quadrupole
moments.

The history of microscopic calculations of quadrupole moments is rather old but, unfortunately, not
rich. Evidently, the first such calculations were made by Bunatyan and Mikulinskii \cite{Bun-Mik} and
by Belyakov \cite{Bel}, within the initial version of the TFFS \cite{AB}, which is not
self-consistemt. In the first of  these references, quadrupole moments were studied together with the
isotopic shifts of atomic levels which are related directly to the isopic variation of charge radii.
This work played an important role in the development of the TFFS, as, for the first time, a strong
density dependence of the scalar-isoscalar Landau--Migdal (LM) interaction amplitude was proposed to
describe both the phenomena simultaneously. Such the density dependence was accepted in the canonical
TFFS theory \cite{AB}.

The first self-consistent  consideration of  the quadrupole moments was carried out only recently in
the Refs. \cite{BE2,QEPJ,QEPJ-Web} cited above. The only alternative modern calculations we know
concern the medium atomic weight nuclei with $A<90$ \cite{SM1,SM2}. They are carried out within the
many-particle Shell Model. This approach is rather comprehensive as it takes into account various
many-particle correlations. However, it is not self-consistent, i.e. the nuclear potential well and
the effective interaction are parameterized independently. In the result, it contains rather big
amount of free parameters, the effective proton and neutron effective charges being among them, which
are fitted for any nuclear shell anew. For nuclei which we consider, with $A>100$, this approach was
not yet applied, evidently, because of technical difficulties.

The self-consistent TFFS uses the energy density functional (EDF) method by Fayans \cite{Fay1,Fay} to
generate the nuclear potential well and the effective $NN$ force as well.  In the modern
self-consistent nuclear theory based on the EDF method, a trend appeared last decade to go beyond the
mean field approximation. In spherical nuclei, it is usually described in terms of the PC effects.
Direct consideration of the PC corrections within the EDF method is a rather delicate problem as they
are included, on average, to the EDF parameters. Thus, only a fluctuating part of such corrections to
nuclear characteristics should be taken into account, if we do not want to readjust the EDF
parameters.

Up to now, the self-consistent consideration of the PC corrections was limited mainly with such an
object as single-particle energies (SPEs) in magic nuclei
\cite{Litv-Ring,Bort,Dobaczewski,Levels,Baldo-PC}. Different kinds of the self-consistent theories
were used  in the references cited above: the relativistic mean-field theory \cite{RMF}  in
\cite{Litv-Ring}, the Skyrme-Hartree-Fock method \cite{SHF} with various Skyrme EDFs in
\cite{Bort,Dobaczewski,Baldo-PC} and the Fayans EDF method \cite{Fay1,Fay} in \cite{Levels}. In all
these cases, the PC corrections to the SPEs turned out to be rather moderate and it seemed reasonable
to  keep the parameters of the corresponding EDFs unchanged. It is also worth mentioning  more recent
considerations of the one-nucleon spectroscopic factors in semi-magic nuclei \cite{spectr1,spectr2}
and the PC corrections to the {\it ab initio} double mass differences in magic nuclei
\cite{DMD1,DMD2,DMD3}.

Recently, a model was developed  \cite{EPL_mu,Lett_mu,YAF_mu} to find the PC corrections to magnetic
moments of odd semi-magic nuclei  within the self-consistent TFFS \cite{scTFFS}. This approach is
based on the general principles of the TFFS \cite{AB} supplemented with the TFFS self-consistency
relation \cite{Fay-Khod}. The self-consistent basis DF3-a \cite{DF3-a} was used, which is just a small
modification of the initial Fayans EDF DF3 \cite{Fay}. The modification concerns the spin-orbit and
effective tensor terms of the EDF only. The main idea of this model was to separate and explicitly
consider  such PC diagrams that  behave in a non-regular way, depending significantly on the nucleus
and the single-particle state of the odd nucleon. The rest (and the major part) of the PC corrections
is supposed to be regular and included in the initial EDF parameters. In this model, we deal with
semi-magic nuclei, which contain two subsystems with absolutely different properties. Indeed, one of
them is superfluid  whereas the second subsystem is normal. The model is valid for nuclei with odd
nucleon belonging to the non-superfluid subsystem. As it is shown in \cite{EPL_mu}, in this case we
can limit ourselves to the PC corrections to the diagrams in the normal subsystem only. It simplifies
the problem drastically.

In this work we apply this  model to quadrupole moments of odd semi-magic nuclei. We consider the
odd-proton neighbors of the even tin nuclei, i.e. the odd isotopes of In and Sb. Their quadrupole
moments were found previously within the self-consistent TFFS at  the mean-field level in
\cite{BE2,QEPJ,QEPJ-Web}. Only the lowest $2^+$ phonon, which plays the main role in the PC
corrections in these nuclei, is taken into account. One of the PC corrections contains the quadrupole
moment of the phonon under consideration. The quadrupole moments of the $2^+_1$ states in even Sn
isotopes were found by us previously in a self-consistent way with the use of the DF3-a EDF in
\cite{QPRC}. In the case of the magnetic moments, the Bohr--Motelson model \cite{BM2} prescription for
the phonon gyromagnetic ratio, $\gamma_{\rm ph}=Z/A$, was used in \cite{EPL_mu,YAF_mu} to find the
phonon magnetic moment. Another difference between these two problems is in the choice of the
effective interaction entering the equation for the effective field \cite{AB}. In the case of magnetic
moments, this is the spin-dependent Landau--Migdal (LM) interaction amplitude with the set of
parameters additional to those of the EDF. Now, we deal with the scalar LM amplitude which is defined
completely with the EDF we use. Thus, the method we apply is self-consistent completely, and no
parameters additional to those of the EDF are used.

Section 2 contains a brief description of the method we use to find the PC corrections to quadrupole
moments of odd semi-magic nuclei on the base of the model \cite{EPL_mu}. Main general formulas we use
can be obtained from those in \cite{EPL_mu} with the change of the external field angular momentum
$J=1$ for the $M1$ field in \cite{EPL_mu} to $J=2$ for the $E2$ field. Their explicit form for the
quadrupole case may be found in \cite{Q-arXiv}. Section 3 contains the calculation results for In and
Sb chains. Section 4 contains our conclusions.

\section{ Brief formalism for the self-consistent PC corrections to quadrupole moments}

Within the TFFS \cite{AB}, quadrupole moments of odd nuclei are determined in terms of the diadonal
matrix elements \beq Q_{\lambda}= \langle\lambda| V |\lambda\rangle_{m=j},\label{Q_lam}\eeq
$|\lambda\rangle$ being the state of the odd nucleon, of the normal component  $V$ of the effective
field. In superfluid nuclei, we deal, the latter obeys the QRPA-like equation, which can be written in
a compact form as \beq {\hat V}(\omega)={\hat e}_q{\hat V}_0(\omega)+{\hat {\cal F}} {\hat A}(\omega)
{\hat V}(\omega), \label{Vef_s} \eeq where all the terms  are  matrices. In the standard TFFS
notation, we have: \beq {\hat V}=\left(\begin{array}{c}V
\\d_1\\d_2\end{array}\right)\,,\quad{\hat
V}_0=\left(\begin{array}{c}e_q V_0
\\0\\0\end{array}\right)\,,
\label{Vs} \eeq \beq {\hat {\cal F}}=\left(\begin{array}{ccc}
{\cal F} &{\cal F}^{\omega \xi}&{\cal F}^{\omega \xi}\\
{\cal F}^{\xi \omega }&{\cal F}^\xi  &{\cal F}^{\xi \omega }\\
{\cal F}^{\xi \omega }&{\cal F}^{\xi \omega }& {\cal F}^\xi \end{array}\right). \label{Fs} \eeq

The matrix ${\hat A}$ consists of $3\times 3$  integrals over $\eps$ of the products of different
combinations of the Green function $G(\eps)$ and two Gor'kov functios $F^{(1)}(\eps)$ and
$F^{(2)}(\eps)$. They can be found, e. g., in \cite{AB}.

Elements of the matrix (\ref{Fs}) are different variation derivatives of the Fayans EDF depending on
the normal $\rho$ and anomalous $\nu$ densities: \beq E_0=\int {\cal E}[\rho({\bf r}),\nu({\bf r})]
d^3r.\label{E0} \eeq They are: \beq {\cal F}=\frac {\delta^2 {\cal E}}{\delta \rho^2};\;{\cal
F}^{\omega \xi}= {\cal F}^{\xi \omega}=\frac {\delta^2 {\cal E}}{\delta \rho \delta \nu};\; {\cal
F}^{\xi}=\frac {\delta^2 {\cal E}}{\delta \nu^2}. \label{LM} \eeq  Here, ${\cal F}$ is the usual
Landau--Migdal (LM) amplitude, whereas ${\cal F}^{\xi}$ is the effective pairing interaction entering
the gap equation. Isotopic indices in Eqs. (\ref{Vef_s}-\ref{LM}) are omitted for brevity.

For the quadrupole moment problem  we deal, one has \cite{BM1} \beq V_0({\bf r}){=}\sqrt{16\pi/5}r^2
Y_{20}({\bf n})(1+\tau_3)/2. \label{Q0}\eeq The local charges $e_q$ are trivial in this case:
$e_q^p=1,\;e_q^n=0$, due to the Ward indentity \cite{AB}. The equation for the vertex $g_L(\bf r)$
obeys the  homogeneous  equation, corresponding to Eq. (\ref{Q_lam}): \beq {\hat g_L}(\omega)={\hat
{\cal F}}  {\hat A}(\omega) {\hat g_L}(\omega) \label{g_L}. \eeq It is worth to mention that the same
components of the LM amplitude (\ref{LM}) enter to Eqs. (\ref{Q_lam}) and (\ref{g_L}), in contrast to
the case of magnetic  moments \cite{EPL_mu}, where the spin-dependent components of the LM amplitude
remain in Eq. (\ref{Q_lam}) for the effective field. The EDF (\ref{E0}) does not contain such terms,
and the corresponding parameters are additional to those of the EDF under consideration.

\begin{figure}
\centerline {\includegraphics [width=80mm]{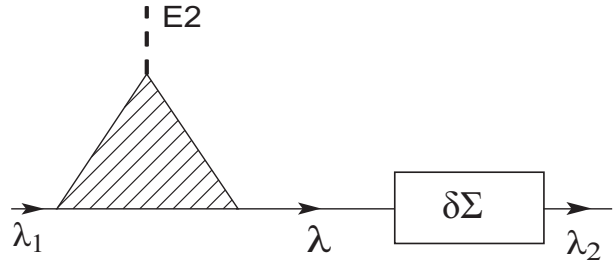}} \vspace{-3mm}\caption{ A diagram of the ``end''
correction. The gray triangle is the effective field $V$ without PC corrections, and the horizontal
slab denotes the variation $\delta \Sigma$ of the mass operator induced by the $L$-phonon.}
\label{fig.1}
\end{figure}

\begin{figure}
\centerline {\includegraphics [width=80mm]{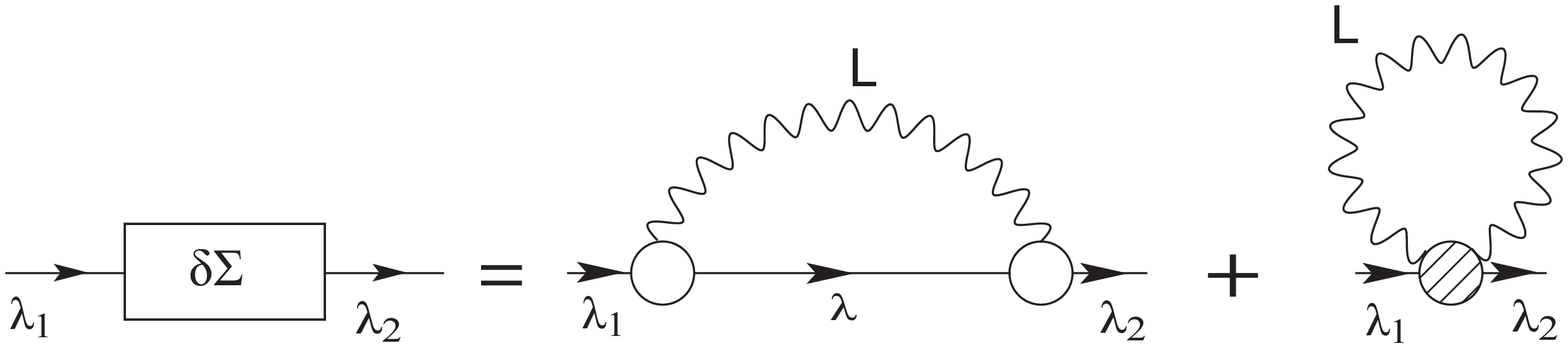}} \vspace{-2mm} \caption{ PC corrections to the mass
operator in the field of the $L$-phonon. The open blob is the $L$-phonon creation amplitude $g_L({\bf
r})$, and the wavy line denotes the phonon $D$-function. The dashed blob denotes the ``tadpole'' term
$\delta\Sigma^{\rm tad}$.} \label{fig.2}
\end{figure}

As it was discussed in the Introduction, for finding the PC corrections to Eq. (\ref{Q_lam}) we use
the self-consistent model developed in \cite{EPL_mu} for the case of magnetic moments. All the
formulas of this article, with only several exceptions, are valid for the case of quadrupole moments
we consider with rather obvious substitution of the $J=2$ for the total angular momentum of the $E2$
field instead of $J=1$ in \cite{EPL_mu}, corresponding to the $M1$ field. The explicit form of these
formulas for quadrupole moments can be found in \cite{Q-arXiv}. Before giving the final formula we
use, let us describe in short how the PC corrections to the matrix element (\ref{Q_lam}) in the field
of the $L$-phonon, can be found: $|\lambda\rangle \to \tilde{|\lambda\rangle}$ and $V\to \tilde{V}$,
with obvious notation. In the general case, when several $L$-phonons are considered, the sum of the
corresponding PC corrections should be found. In this work, we limit ourselves with the $2^+_1$
phonon, which plays the main role in the problem we consider. Following to the scheme of
\cite{EPL_mu}, we limit ourselves to the $g^2_L$-approximation. We consider the ground state of the
odd nucleus without phonons, therefore the first order corrections $\delta^{(1)}_L$ to any observable
vanish. Thus, the second order variations $\delta^{(2)}_L$ of each element of Eq. (\ref{Q_lam})
appear, and mixed $\delta^{(1)}_L$ variations of two different elements of this formula contribute
also. The main idea of the model developed in \cite{EPL_mu} is to separate such PC corrections which
behave in a non-regular way depending significantly on the phonon excitation energy $\omega_L$ and on
the state $|\lambda\rangle$ under consideration. From this point of view, as the analysis of
\cite{EPL_mu} shows, the mixed term with $\delta^{(1)}_L |\lambda\rangle$ and $\delta^{(1)}_L V$ can
be omitted, whereas those containing two corrections $\delta^{(1)}_L |\lambda\rangle$ of different
states $\lambda$ in (\ref{Q_lam}) are very important.

We name the term of $\delta^{(2)}_L |\lambda\rangle$ as the ``end correction''. The corresponding
diagram is displayed in Fig. 1. Obviously, there is a symmetric counterpart where the left end is PC
corrected. Two diagrams for the PC correction to the mass operator $\Sigma(\eps)$ are displayed in
Fig.2. In addition to the usual pole diagrams, the tadpole term is taken into account in our
calculations which is very important to find the total PC correction to the single-particle energy
$\eps_{\lambda}$ \cite{Levels}. Two situations are possible in the diagram of Fig. 1. The first one,
when we have $\lambda=\lambda_2$, or $\lambda=\lambda_1$ for the PC correction to the left end. The
corresponding terms in the end correction expression, see Eq. (8) in \cite{EPL_mu},  are singular.
This singularity is removed with the standard renormalization \cite{scTFFS} of the single particle
wave functions : $|\lambda\rangle \to \sqrt{Z_{\lambda}} |\lambda\rangle$, where \beq Z_{\lambda} =
\left( 1- \left. \frac {\partial \delta \Sigma_{\lambda\lambda}(\eps)}{\partial
\eps}\right|_{\eps=\eps_{\lambda}}\right)^{-1} \label{Zlam} \eeq is the residue of the Green function
at the pole $\eps=\eps_{\lambda}$. It is seen that the expression (\ref{Zlam}) is going beyond the
perturbation $g_L^2$ approximation, corresponding to a partial summing of the chain of the pole
diagrams in Fig. 2. The corresponding perturbation analog is \beq Z_{\lambda}^{\rm ptb} =   1+ \left.
\frac {\partial \delta \Sigma_{\lambda\lambda}(\eps)}{\partial \eps}\right|_{\eps=\eps_{\lambda}}
\label{Z_ptb} \eeq   The second situation takes place for the terms with $\lambda \neq \lambda_{1,2}$,
which are regular.  In accordance with \cite{EPL_mu}, we denote the corresponding correction to the
effective field $V$ as $\delta V'_{\rm end}$.

\begin{figure}
\centerline {\includegraphics [width=40mm]{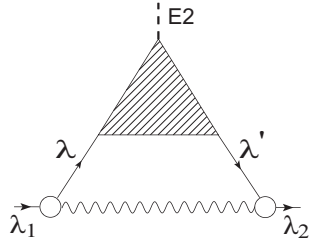}} \vspace{-3mm}\caption{ The induced interaction
correction.} \label{fig.3}
\end{figure}

The term which involves two corrections $\delta^{(1)}_L |\lambda\rangle_1$ and  $\delta^{(1)}_L
|\lambda\rangle_2$ is displayed in Fig. 3. As it is seen, it corresponds to the phonon-induced
interaction. In accordance with \cite{EPL_mu}, the corresponding PC correction to $V$ is denoted as
$\delta V_{GGD}$, as it involves the integral of two particle Green functions $G$ and one phonon
$D$-function. At last,  consider the term  which corresponds to the direct action of the external
field to the $L$-phonon under consideration. It is displayed in Fig. 4, and the corresponding PC
correction to $V$ is denoted as $\delta V_{GDD}$, as it involves the integral of one Green functions
$G$ and two phonon $D$-functions. In Eq. (\ref{Vef_s}), it appears due to the variation of the term of
$e_q$. The black blob denotes the quadrupole moment of the $L$-phonon. For the analogous quantity for
magnetic moments, \cite{EPL_mu} the Bohr-Mottelson model approximation for the phonon gyromagnetic
ratio was used, $\gamma^{\rm ph}=Z/A$. Unfortunately, there is no similar simple approximation for the
phonon quadrupole moments. The corresponding values for the $2^+$-phonons we consider were found by us
previously \cite{QPRC} in a self-consistent way.

As it is shown in \cite{EPL_mu}, this term consists of two components, $\delta V_{GDD}=\delta
V_{GDD}^{(1)}+\delta V_{GDD}^{(2)}$, with different behavior at $\omega_L \to 0$. The first term is
regular at small $\omega_L$, see Eq. (21) in \cite{EPL_mu}, whereas the second one behaves as
$1/\omega_L$, Eq. (21) in \cite{EPL_mu}. There \textbf{is} also a tadpole-like counterpart to Fig. 4,
which is displayed in Fig. 5. As it is shown in \cite{EPL_mu}, it also  behaves  as $1/\omega_L$ at
small $\omega_L$, possessing the opposite sign  compared  to the term of $V_{GDD}^{(2)}$. In the model
developed in \cite{EPL_mu} it is supposed that these two divergent $\omega_L\to 0$ terms cancel each
other exactly.

\begin{figure}
\centerline {\includegraphics [width=40mm]{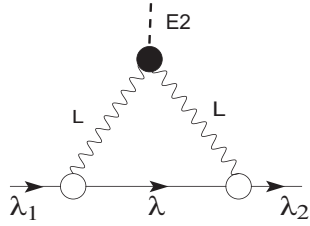}} \vspace{-3mm}\caption{Correction due to the direct action of
the $E2$-field to the phonon. The black blob is the quadrupole moment of the phonon.} \label{fig.4}
\end{figure}

\begin{figure}[]
\centerline {\includegraphics [width=15mm]{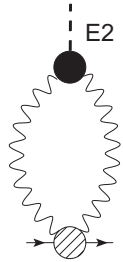}} \vspace{-3mm} \caption{ The
tadpole-like diagram for the contribution of the phonon quadrupole moment.}
\end{figure}

  \begin{table}[h!]
\caption{Characteristics of the $2^+_1$ phonons in even Sn isotopes: excitation energies $\omega_2$
(MeV) and quadrupole moments $Q(2^+_1)$(b)}
\begin{indented}
\item[]\begin{tabular}{|c| c| c| c| c|  }

\hline\noalign{\smallskip} $A$  & $\omega_2^{\rm th}$   & $\omega_2^{\rm exp}$
&  $Q^{\rm th}$ &  $Q^{\rm exp}$  \\
\noalign{\smallskip}\hline\noalign{\smallskip}
106    & 1.316      & 1.207    & -0.34    &   -    \\
108    & 1.231      & 1.206    & -0.39    &       \\
110    & 1.162      & 1.212    & -0.50    &       \\
112    & 1.130      & 1.257    & -0.45    & -0.03(11)    \\
114    & 1.156      & 1.300    & -0.28    &  0.32(3)        \\
116    & 1.186      & 1.294    & -0.12    & -0.17(4)      \\
118    & 1.217      & 1.230    & -0.01    & -0.05(14)      \\
120    & 1.240      & 1.171    &  0.04    & +0.022(10)      \\
122    & 1.290      & 1.141    &  0.01    & $< +0.14$    \\
124    & 1.350      & 1.132    & -0.07    &  0.0(2)    \\
126    & 1.405      & 1.141    & -0.13   &       \\
128    & 1.485      & 1.169    & -0.14    &       \\

\noalign{\smallskip}\hline

\end{tabular}\label{tab1}
\end{indented}
\end{table}

The final ansatz for the quadrupole moments with PC corrections is as follows: \beq
\widetilde{V}_{\lambda\lambda} {=} Z_{\lambda} \left( V + \delta V_{GGD} + \delta V^{(1)}_{GDD} +
\delta V'_{\rm end}\right)_{\lambda\lambda}. \label{final} \eeq  It corresponds to making ``fat'' the
ends in Figs. 3--5: $|\lambda\rangle \to \sqrt{Z_{\lambda}} \; |\lambda\rangle$. In Refs.
\cite{EPL_mu,YAF_mu} it was used for magnetic moments of odd semi-magic nuclei. Here we apply this
model to the quadrupole moments of odd In and Sb isotopes.

\section{ PC corrected quadrupole moments of In and Sb isotopes}

The self-consistent scheme of solving Eqs. (\ref{Vef_s}) and (\ref{g_L}) is described in detail in
\cite{BE2} and \cite{QEPJ}. The DF3-a version \cite{DF3-a} of the Fayans EDF is used, which differs
from the initial EDF DF3 \cite{Fay} with parameters of the spin-dependent terms of the EDF (\ref{E0}),
the spin-orbit and effective tensor ones. The excitation energies and quadrupole moments of $2^+_1$
states in even Sn isotopes, which are ingredients of all equations of \cite{EPL_mu}, are presented in
Table 1. They are taken from \cite{BE2} and \cite{QPRC}, correspondingly. In general, the agreement
with data looks reasonable, however, the accuracy of reproducing experimental excitation energies for
heavy tin isotopes, $A>120$, is worse than for lighter ones. In Ref. \cite{BE2} a sensitivity of the
$\omega_2$ values to the spin-dependent terms of the EDF (\ref{E0}). In particular, the DF3-a EDF
proved out to be better than the DF3 one. However, the accuracy of both Fayans EDFs is significantly
higher than of all the predictions of Skyrme EDFs we know \cite{BE2-HFB}.

Table 2 represents different PC corrections to quadrupole moments discussed above. Here, the
$Z$-factor, column 4, is found from its definition (\ref{Zlam}), whereas the perturbation theory
prescription (\ref{Z_ptb}) is used for finding the $\delta Q^Z_{\rm ptb}$ values in column 5. Thus,
the $\delta Q_{\rm ptb}$ quantity is just the sum of four partial corrections in previous columns. At
last, the  quantity in the last column is $\delta Q_{\rm ph}=\tilde{V}-V$ where $\tilde{V}$ is the
result of the use of Eq. (\ref{final}) which is the final prescription of the model we use. We see
that two main corrections are those due to the $Z$-factor (column 5) and due to the induced
interaction (the term $\delta Q_{GGD}$, column 6). They always possess different signs, the sum being
significantly less in the absolute value than each of them. Therefore two other ``small'' corrections
are sometimes also important. It is worth to mention that the $Z$-factor values are often about 0.5
which makes the use of the perturbation theory in  $g_L^2$ questionable as the value of $(1-Z)$ is a
measure of validity of the  $g_L^2$ approximation. Eq. (\ref{final}) we use contains higher in $g_L^2$
terms, but it is just an ansatz. The analysis shows that the $g_L^2$ approximation in semi-magic
nuclei is valid on the average, but often one ``dangerous'' term appears in Eq. (8) of \cite{EPL_mu}
with small energy denominator leading to a big contribution to the value of $(1-Z)$. Simultaneously,
the same small denominator contributes, with opposite sign, to Eq. (21) of \cite{EPL_mu} for the
induced interaction correction. In the result, two inaccuracies compensate each other partially.
However, a more consistent approach should be developed for the small denominator situation. Such
development is in our nearest plans. The first step is made, for the problem of the PC corrections to
the single-particle energies of semi-magic nuclei, in \cite{semiLevs}, where the Dyson equation with
the PC-corrected mass operator is solved directly, without any use of PT.

\begin{table*}
\caption{Different PC corrections to the quadrupole moments of odd-proton In and Sb nuclei. $Q$ is the
quadrupole moment without PC corrections \cite{QEPJ}. Other notation is explained in the text. All
values, except $Z$, are in b.}
\begin{indented}
\item[]\begin{tabular}{|l| c| c| c| c| c| c| c| c| c|}
\noalign{\smallskip}\hline\noalign{\smallskip}    nucl.  &$\lambda$ & $Q$ & $Z$ &$\delta Q^Z_{\rm
ptb}$& $\delta Q_{GGD}$ & $\delta Q_{GDD}$ &
$\delta Q_{\rm end}'$ & $\delta Q_{\rm ptb} $  & $\delta Q_{\rm ph} $\\
\noalign{\smallskip}\hline\noalign{\smallskip}
$^{105}$In & $1g_{9/2}$ &  +0.833 &0.675 &-0.400 &0.231 &0.055 & 0.014 &-0.100  &-0.067\\

$^{107}$In & $1g_{9/2}$ &  +0.976 &0.584 &-0.692 &0.404 &0.094 & 0.021 &-0.172  &-0.100\\

$^{109}$In & $1g_{9/2}$ & +1.113 &0.573 &-0.826 &0.487 &0.128 & 0.023 &-0.188 &-0.108\\

$^{111}$In & $1g_{9/2}$ & +1.165 &0.488 &-1.220&0.722 &0.163  & 0.034 &-0.301 &-0.147\\

$^{113}$In & $1g_{9/2}$ & +1.117 &0.576 &-0.820 &0.484 &0.071 & 0.025 &-0.240 &-0.138\\

$^{115}$In & $1g_{9/2}$ & +1.034 &0.609 &-0.662 &0.389 &0.026 &0.023 &-0.224 &-0.136\\

$^{117}$In & $1g_{9/2}$&   +0.963 &0.632 &-0.560 &0.328 &0.002 &0.021 &-0.209 &-0.132  \\

$^{119}$In & $1g_{9/2}$&   +0.909 &0.621 &-0.553 &0.323 &-0.008&0.022 &-0.216 &-0.134 \\

$^{121}$In & $1g_{9/2}$&   +0.833 &0.639 &-0.465 &0.271 &-0.002&0.019 &-0.177 &-0.113 \\

$^{123}$In & $1g_{9/2}$&   +0.743 &0.720 &-0.289 &0.168 &0.009 &0.011 &-0.101 &-0.073  \\

$^{125}$In & $1g_{9/2}$&   +0.663 &0.738 &-0.232 &0.134 &0.015 &0.010 &-0.073  &-0.054 \\

$^{127}$In & $1g_{9/2}$&   +0.550 &0.800 &-0.138 & 0.079  &0.012 &0.006 &-0.041  &-0.033  \\

$^{115}$Sb & $2d_{5/2}$&  -0.882  &0.551 &0.717  &-0.275 &-0.025   &0.053 &0.470  &0.259  \\

$^{117}$Sb & $2d_{5/2}$&  -0.817  &0.582 &0.588  &-0.229 &-0.009    &0.050 &0.399 &0.232 \\

$^{119}$Sb & $2d_{5/2}$&  -0.763  &0.602 &0.504  &-0.198 &-0.001 &0.048   &0.353 &0.213 \\

$^{121}$Sb & $2d_{5/2}$&  -0.721  &0.591 &0.497  &-0.196 &0.003  &0.052   &0.355 &0.210 \\

$^{123}$Sb  & $1g_{7/2}$& -0.739  &0.570 &0.552  &-0.328 &0.001  &0.099 &0.323 &0.184 \\

\noalign{\smallskip}\hline\noalign{\smallskip}
\end{tabular}
\label{tab:delQ_ph}
\end{indented}
\end{table*}

\begin{table*}
\begin{center} \caption{Quadrupole moments $Q\;$(b) of odd In and Sb isotopes. Experimental data are
taken from the review article \cite{Stone-2014}. For the $^{115}$In isotope, the first value
corresponds to the original experiment of \cite{exp-In115-1}, the second one, to \cite{exp-In115-2}.
Similarly, for the $^{121}$Sb isotope, the first value corresponds to \cite{exp-Sb121-1}, the second
one, to \cite{exp-Sb121-2}. $Q_0$ is the prediction of the single-particle model. The theoretical
values are $Q_{\rm th}$ and $\tilde{Q}_{\rm th}$ without and with PC corrections, correspondingly. The
differences $\delta Q= Q_{\rm th}-Q_{\rm exp}$ and $\delta \tilde{Q}=\tilde{Q}_{\rm th}-Q_{\rm exp}$
are given in the last two columns.}

\begin{tabular}{|l |c |c |c |c |c |c|c|}
\noalign{\smallskip}\hline\noalign{\smallskip}
  nucl.  &$\lambda$  & $Q_{\rm exp}$& $Q_0$ &
$Q_{\rm th}$ &$\tilde{Q}_{\rm th}$& $\delta Q$ &$\delta \tilde{Q}$\\
\noalign{\smallskip}\hline\noalign{\smallskip}

$^{105}$In & $1g_{9/2}$& +0.83(5)&0.18 & +0.83 & 0.76 &0.00 &-0.07 \\

$^{107}$In & $1g_{9/2}$& +0.81(5) &0.18& +0.98 &  0.87&0.17 &  0.06   \\

$^{109}$In & $1g_{9/2}$& +0.84(3) &0.18& +1.11 &  1.00&0.27 &  0.16\\

$^{111}$In & $1g_{9/2}$& +0.80(2) &0.19 &+1.17 &  1.02&0.37&  0.22\\

$^{113}$In & $1g_{9/2}$& +0.80(4) &0.19& +1.12 &  0.98&0.32 &  0.16\\

$^{115}$In & $1g_{9/2}$& +0.81(5) &0.19 &+1.03 &  0.90&0.22 &  0.09 \\

&                       & 0.58(9)&   &&           &0.45 & 0.32\\

$^{117}$In & $1g_{9/2}$& +0.829(10)&0.19& +0.96   & 0.83 &0.131 &0.001  \\

$^{119}$In & $1g_{9/2}$& +0.854(7) &0.19 &+0.91   & 0.773&0.056 &-0.081\\

$^{121}$In & $1g_{9/2}$& +0.814(11) &0.19 &+0.833 & 0.711&0.019 &-0.103 \\

$^{123}$In & $1g_{9/2}$& +0.757(9)  &0.19 &+0.743 & 0.670&-0.014 &-0.087 \\

$^{125}$In & $1g_{9/2}$& +0.71(4)   &0.19 &+0.66  & 0.60 &-0.05 &-0.11 \\

$^{127}$In & $1g_{9/2}$& +0.59(3)   &0.19 &+0.55  & 0.52 &-0.04 &-0.07  \\

$^{115}$Sb & $2d_{5/2}$& -0.36(6)   &-0.14 &-0.88  & -0.62&-0.52 &-0.26  \\

$^{117}$Sb & $2d_{5/2}$&    -       &-0.14 &-0.817 & -0.585&- &-   \\

$^{119}$Sb & $2d_{5/2}$& -0.37(6)   &-0.14 &-0.76  & -0.55 &-0.39 &-0.18 \\

$^{121}$Sb & $2d_{5/2}$& -0.36(4)   &-0.14&-0.72  & -0.51 &-0.36 &-0.15 \\
&                      & -0.45(3)   & &      &       &-0.27 &-0.06 \\

$^{123}$Sb  & $1g_{7/2}$& -0.49(5)  &-0.17 &-0.74 & -0.55 &-0.25 &-0.06 \\

\hline
\end{tabular}
\end{center}
\label{tab:Q_p}
\end{table*}

\begin{figure}[h!]
\centerline {\includegraphics [width=120mm]{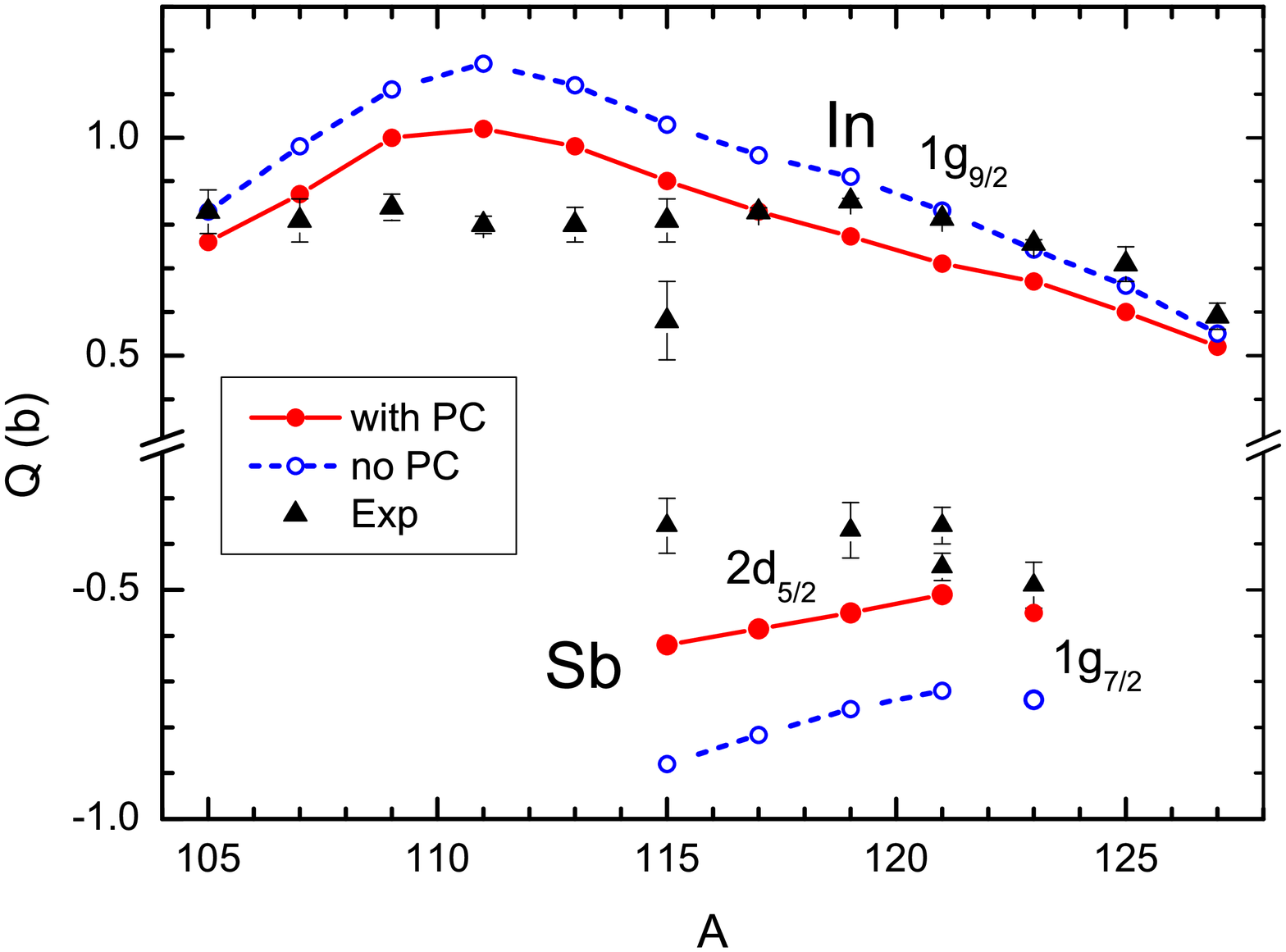}}  \caption{Quadrupole moments of odd In and Sb
isotopes with and without PC corrections. Experimental data are taken from \cite{Stone-2014}. For the
$^{115}$In isotope, the upper value corresponds to the original experiment of \cite{exp-In115-1}, the
lower one, to \cite{exp-In115-2}. Similarly, for the $^{121}$Sb isotope, the upper value corresponds
to \cite{exp-Sb121-1}, the lower one, to \cite{exp-Sb121-2}.}
\end{figure}

The final results are presented in Table 3 and Fig. 6. Experimental data are taken from the review
article \cite{Stone-2014}. In the case if there are several different experimental values, that
overlap with their bars, we, as a rule, choose the one with minimal errors. There are two cases,
$^{115}$In and $^{121}$Sb isotopes, when the two values do not overlap. In  these cases, we present
both the values giving the references to the original experimental works. To show the scale of the
effects we describe, the single-particle values $Q_0$ of quadrupole moments are also presented. They
are found as matrix elements of (\ref{Q0}) multiplied by the Bogolyubov factor of $u^2_{\lambda} -
v^2_{\lambda}$ \cite{Sol}, with obvious notation. They always possess correct signs, but the absolute
values are to 3 - 5 times less than the experimental data. We see that the PC corrections to
quadrupole moments taken into account make agreement with experiment better in most cases. In any
case, this is true in all the cases where the deviation from experiment of the results without PC
corrections is significant, more than 0.1 b, which is evidently a typical accuracy of the theory we
develop. For example, this is so for nuclei $^{109,111}$In and for all Sb isotopes. The rms value
$<\delta \tilde{Q}>_{\rm rms}=0.15\;$b follows from the last column of Table 3. The corresponding
value without PC corrections is significantly bigger, $<\delta {Q}>_{\rm rms}=0.27\;$b. Note that the
value of \cite{exp-In115-2} for $^{115}$In which gives the maximum value of the deviation $\delta
\tilde{Q}$ of the PC-corrected theory from experiment below from the general tendency of the $Q_{\rm
exp}$ values, as is seen in Fig. 6. We think that the value of \cite{exp-In115-1} is closer to the
truth. Note also that, for the In chain, the $(1-Z)$ difference, which is a measure of validity of the
PT, is maximal for the $^{111}$In isotope and its neighbors for which the error of our theory is also
maximal. We hope that a more consistent theory which is in our plans will result in better agreement
with experiment.

\newpage
\section{Conclusions}

We have developed a model of the self-consistent account of the PC corrections to quadrupole moments
of odd semi-magic nuclei for the case where the odd nucleon belongs to the non-superluid component of
the nucleus under consideration. The main ingredients of the calculation scheme are similar to those
proposed recently by us for magnetic moments \cite{EPL_mu,YAF_mu}.  The perturbation theory in $g_L^2$
is used, $g_L$ being the vertex of creating the $L$-phonon. The main idea of our approach is to refuse
from calculation of all terms proportional to $g_L^2$, as their main  part is taken into account
implicitly in the EDF parameters we use. Instead, only such $g_L^2$ diagrams are separated and
calculated explicitly which contribution fluctuates, depending on the nucleus under consideration and
the state $\lambda$ of the odd nucleon. The term $\delta Q^Z$ due to the phonon $Z$-factor and the
one, $\delta Q_{GGD}$, due to the phonon-induced interaction are the two main such terms. However,
they always possess opposite signs and cancel each other significantly. Therefore two ``small''
corrections, the one, $\delta Q_{GDD}$, due to the phonon quadrupole moment and the non-diagonal ``end
term'' $\delta Q_{\rm end}'$ are also often important. The sum of all four PC corrections to
quadrupole moments of odd In and Sb isotopes, in most cases, improves the agreement with experiment.
In any case, this is so always when the deviation of the mean field prediction for the quadrupole
moment from the experimental value is significant, more than 0.1 b. For the sample of 18 nuclei we
consider, the rms value of the difference between the theoretical predictions and experimental values
is now $<\delta \tilde{Q}>_{\rm rms}=0.15\;$b, instead of the value of $<\delta {Q}>_{\rm
rms}=0.27\;$b for the calculation without PC corrections.

The difference of $(1-Z)$ is a measure of validity of the $g_L^2$ approximation. In nuclei we consider
there are several cases with $Z\simeq 0.5$, which makes application of the perturbation theory in
$g_L^2$ questionable. As it was discussed above, this occurs, as a rule, due to appearance of a
``dangerous'' term with small energy denominator in Eq. (7) in \cite{EPL_mu} for the PC correction to
the mass operator $\Sigma$, and the analogous term arises simultaneously, in the correction $\delta
Q_{GGD}$ of the phonon-induced interaction. In the sum of these two terms, the inaccuracy becomes
smaller then for each term separately which partially saves the approximation we use. However, a more
consistent consideration is necessary of the dangerous cases, and this is in our nearest plans. The
first step is made in \cite{semiLevs}, where the method is developed to find the PC corrections to
SPEs in semi-magic nuclei beyond the perturbation theory.

\section{Acknowledgments} The work is supported with the Russian Science Foundation,  Grants Nos.
16-12-10155 (Section 3) and 16-12-10161 (Section 2).
 We thank O. I. Achakovskiy  for help. Calculations were partially carried out on the
Computer Center of NRC ``Kurchatov Institute''. EES was partially supported by the Academic Excellence
Project of the NRNU MEPhI under contract with the Ministry of Education and Science of the Russian
Federation No. 02.A03.21.0005.

\end{document}